\documentclass[preprint,showpacs,preprintnumbers,amsmath,amssymb]{revtex4}

\usepackage{graphicx}
\usepackage{dcolumn}
\usepackage{bm}

\begin{document}

\preprint{}

\title{Univariate Functions for the Hilbert-Schmidt 
Volumes of
the Real and Complex Separable Two-Qubit Systems}

\author{Paul B. Slater}%
\email{slater@kitp.ucsb.edu}
\affiliation{%
ISBER, University of California, Santa Barbara, CA 93106\\
}%
\date{\today}

\begin{abstract}
The (complex) two-qubit systems comprise a 15-dimensional convex set and
the real two-qubit systems, a 9-dimensional convex set. While formulas
for the Hilbert-Schmidt volumes of these two sets are known --- owing to
recent important work of Sommers and \.Zyczkowski 
({\it J. Phys. A} {\bf{36}}, 10115 [2003]) --- formulas
 have not been so far obtained for the volumes of the 
{\it separable}  subsets.  We reduce these two
problems to the determination of certain  functions ($f_{real}(\mu)$ 
and $f_{complex}(\mu)$)
of a {\it single} variable $\mu = 
\sqrt{\frac{\rho_{11} \rho_{44}}{\rho_{22} \rho_{33}}}$, where $\rho$
is the corresponding $4 \times 4$ density matrix, and the $\rho_{ii}$'s
($i=1,\ldots4$) its diagonal entries.
The desired separable volumes are, then, of the form
$V_{sep/real}= 2 \int_{0}^{1} jac_{real}(\mu) f_{real}(\mu) d \mu$ and
$V_{complex/real} = 2 \int_{0}^{1} jac_{complex}(\mu) 
f_{complex}(\mu) d \mu$. 
Here $jac$ denotes
the corresponding (known) jacobian function. 
We provide estimates of  the two sets of $f(\mu)$'s and $V$'s.
\end{abstract}


\maketitle
\section{Introduction}
In a pair of major, skillful papers, making use of the theory of
 random matrices \cite{random}, Sommers and \.Zyczkowski were able to
derive explicit formulas for the volumes occupied by the 
$d= (n^2-1)$-dimensional convex set of $n \times n$ (complex) 
density matrices (as well as the $d=\frac{(n-1)(n+2)}{2}$-dimensional 
convex set of real (symmetric) $n \times n$ density matrices),
both in terms of the Hilbert-Schmidt (HS) metric \cite{szHS} --- inducing the flat, Euclidean geometry --- and 
the Bures metric \cite{szBures} (cf. \cite{szMore}).
Of course, it would be of obvious 
considerable quantum-information-theoretic 
interest in the cases that $n$ is a composite
number, to also obtain HS and Bures volume 
formulas restricted to those states that
are separable --- the sum of 
product states --- in terms of some factorization of $n$ \cite{ZHSL}. 
Then, by taking ratios --- employing these Sommers-\.Zyczkowski 
results --- one would obtain corresponding 
separability {\it probabilities}.
(In an {\it asymptotic} regime, in which the dimension of the state space
grows to infinity, Aubrun and Szarek 
recently concluded \cite{aubrun} that for
qubits and larger-dimensional particles, the proportion of the
states that are separable is superexponentially small in the dimension
of the set.)

In particular, again for the 15-dimensional complex 
case, $n=4 = 2 \times 2$, numerical
evidence has been adduced 
that the Bures volume of separable states is
(quite elegantly) $2^{-15} (\frac{\sqrt{2}-1}{3}) \approx 
4.2136  \cdot 10^{-6}$ \cite[Table VI]{slaterJGP} 
and the HS volume
$(5 \sqrt{3})^{-7} 
\approx 2.73707 \cdot 10^{-7}$ \cite[eq. (41)]{slaterPRA}. 
Then, taking ratios (using the corresponding Sommers-\.Zyczkowski results), 
we have the derived conjectures that the Bures separability
 probability
is $\frac{1680 (\sqrt{2}-1)}{\pi^8} \approx 0.0733389$ 
and the HS one, 
considerably larger, $\frac{2^2 \cdot 3 \cdot 7^2 \cdot 11 \cdot 13 \sqrt{3}}{5^4 \pi^{6}} \approx 0.242379$ \cite[eq. (43), but misprinted as $5^3$ not 
$5^4$ there]{slaterPRA}.
(Szarek, Bengtsson and \.Zyczkowski --- motivated by the numerical 
findings of \cite{slaterPRA,slaterChinese} --- have recently 
formally demonstrated 
 ``that the probability to find a random state to be separable equals 2 times the probability to find a random boundary state to be separable, provided the random states are generated uniformly with respect to the Hilbert-Schmidt (Euclidean) distance. An analogous property holds for the set of positive-partial-transpose states for an arbitrary bipartite system'' \cite{sbz} 
(cf. \cite{innami}). These authors 
also noted \cite[p. L125]{sbz} that ``one could try to obtain similar
results for a general class of multipartite systems''. In this 
latter vein, recent 
numerical analyses of ours give some [but certainly not yet conclusive] 
indication that for the {\it three}-qubit
{\it tri}separable states, there is an analogous probability ratio of
6 --- rather than 2.)

However, the analytical derivation of (conjecturally) exact
formulas for these HS and Bures (as well as other, such as the Kubo-Mori
\cite{petz1994} and Wigner-Yanase 
\cite{wigneryanase,slaterPRA}) 
separable volumes has seemed quite remote --- the only analytic 
progress to report so far being 
certain exact formulas
when the number of dimensions of the 15-dimensional space of $4 \times 4$ 
density matrices has been severely 
curtailed (nullifying or holding 
constant {\it most} of the 15 parameters) to $d \leq 3$ 
\cite{pbsJak,pbsCanosa} (cf. \cite{slaterC}).
Most notably, in this research direction,
in \cite[Fig. 11]{pbsCanosa}, we were able to find 
a highly interesting/intricate (one-dimensional) continuum 
($-\infty < \beta <\infty$) of two-dimensional 
(the associated 
parameters being $b_{1}$, the mean, and $\sigma_{q}^2$,   
the variance of the Bell-CHSH observable) 
HS separability
probabilities, in which the {\it golden ratio} \cite{livio} was 
featured, among other items. (The associated 
HS volume element --- $\frac{1}{32 \beta (1+\beta)} 
d \beta d b_{q} d \sigma^2_{q}$ --- is
independent of $b_{1}$ and $\sigma_{q}^2$ in this 
three-dimensional scenario.)
Further, in \cite{pbsJak}, building upon work of 
Jak\'obczyk and Siennicki \cite{jak}, we obtained a 
remarkably wide-ranging variety of exact HS 
separability ($n=4, 6$) and PPT (positive partial transpose) 
($n=8, 9, 10$) probabilities based on 
{\it two}-dimensional sections of sets of 
(generalized) Bloch vectors corresponding to $n \times n$ 
density matrices.

The full $d=9$ and/or $d =15$, $n=4$ real and complex 
two-qubit scenarios
are quite daunting --- due to the numerous
separability constraints at work, some being active [binding] 
in certain regions and 
in complementary regions, inactive [nonbinding]. 
``The geometry of the $15$-dimensional set of separable states of two
qubits is not easy to describe'' \cite[p. L125]{sbz}.
We will seek to make substantial progress in these directions here, 
by recasting  both these problems within a {\it one}-dimensional 
framework.

To proceed in our study,
we employ the (quite simple) form of parameterization of the density matrices
put forth by Bloore \cite{bloore,slaterJPAreject} some thirty years ago. 
(Of course, there are a number of
other possible parametrizations \cite{kk,byrd,sudarshan,vanik,fano,scutaru,stan}, a number of 
which we have also utilized in various studies \cite{slaterA,slaterqip} 
to estimate volumes of 
separable states. Our greatest progress at this stage, 
in terms of increasing dimensionality,  has been achieved
with the Bloore parameterization --- due to a certain 
computationally attractive feature of it, allowing us to 
decouple diagonal and non-diagonal parameters --- as detailed 
shortly below.)

\section{Bloore parameterization of the density matrices} \label{sc1}
The main presentation of Bloore \cite{bloore} 
was made in terms of the $3 \times 3$ ($n=3$)
density matrices. It is clearly easily extendible to cases
$n >  3$.
The fundamental idea is to scale the off-diagonal elements $(\rho_{ij}, 
i \neq j)$ 
of the density matrix in terms of the square roots of the diagonal 
entries ($\rho_{ii}$). That is, we set (introducing the new [Bloore] 
variables $z_{ij}$),
\begin{equation}
\rho_{ij} = \sqrt{\rho_{ii} 
\rho_{jj}} z_{ij}.
\end{equation}
 This allows the determinant of $\rho$ (and analogously 
all its
principal minors) to be expressible as the product
($|\rho| = A B$) of
two factors, one ($A = \Pi_{i=1}^{4} \rho_{ii}$)  
of which is itself simply the product of 
(nonnegative) 
diagonal entries ($\rho_{ii}$). 
In the real $n=4$ case under investigation here --- we have
\begin{equation} \label{B}
B= \left(z_{34}^2-1\right) z_{12}^2+2 \left(z_{14}
   \left(z_{24}-z_{23} z_{34}\right)+z_{13}
   \left(z_{23}-z_{24} z_{34}\right)\right)
   z_{12}-z_{23}^2-z_{24}^2-z_{34}^2+
\end{equation}
\begin{displaymath}
z_{14}^2
   \left(z_{23}^2-1\right)+ z_{13}^2
   \left(z_{24}^2-1\right)+2 z_{23} z_{24} z_{34}+2 z_{13}
   z_{14} \left(z_{34}-z_{23} z_{24}\right)+1,
\end{displaymath}
involving (only) the $z_{ij}$'s ($i > j$), where $z_{ji}=z_{ij}$ 
\cite[eqs. (15), (17)]{bloore}.
Since, clearly, the factor $A$ is positive in all nondegenerate cases 
($\rho_{ii} > 0$),
one can --- by only analyzing $B$ --- essentially 
ignore the diagonal entries, and thus reduce by ($n-1$) the
dimensionality of the problem of finding nonnegativity 
conditions to impose on $\rho$.
This is the feature we will seek to maximally 
exploit here. A fully analogous 
decoupling property holds in the complex case.

It is, of course, necessary and sufficient for $\rho$ to serve
as a density matrix (that is, an Hermitian, nonnegative definite, trace
one matrix) that all its principal minors be nonnegative 
\cite{horn}.
The condition --- quite natural in the Bloore 
parameterization --- that all the principal $2 \times 2$ minors be
nonnegative requires simply that $-1 \leq z_{ij} \leq 1, i \neq j$. The 
joint conditions that all the principal minors be nonnegative are not as
readily apparent. But for the 9-dimensional {\it real} 
case $n=4$ --- that is, $\Im(\rho_{ij})=0$ --- we have been able to obtain 
one such set,
using the Mathematica implementation of the {\it cylindrical
algorithm decomposition} \cite{cylindrical}.
(The set of solutions of any system of real algebraic equations
and inequalities can be decomposed into a finite number of
``cylindrical'' parts \cite{strzebonski}.)
Applying it, we were able to express the 
conditions that 
an arbitrary  9-dimensional $4 \times 4$ real density matrix 
$\rho$ must fulfill.
These took the form, $z_{12}, z_{13}, z_{14} \in [-1,1]$ and 
\begin{equation} \label{limits}
 z_{23} \in [Z^-_{23},Z^+_{23}],
 z_{24} \in [Z^-_{24},Z^+_{24}],
 z_{34} \in [Z^-_{34},Z^+_{34}],
\end{equation}
where
\begin{equation}
Z^{\pm}_{23} =z_{12} z_{13} \pm \sqrt{1-z_{12}^2} \sqrt{1-z_{13}^2} , 
Z^{\pm}_{24} =z_{12} z_{14} \pm \sqrt{1-z_{12}^2} \sqrt{1-z_{14}^2} ,
\end{equation}
\begin{displaymath}
Z^{\pm}_{34} = \frac{z_{13} z_{14} -z_{12} z_{14} z_{23} -z_{12} z_{13} z_{24} +z_{23}
z_{24} \pm s}{1-z_{12}^2},
\end{displaymath}
and
\begin{equation}
s = \sqrt{-1 +z_{12}^2 +z_{13}^2 -2 z_{12} z_{13} z_{23} +z_{23}^2}
\sqrt{-1 +z_{12}^2 +z_{14}^2 -2 z_{12} z_{14} z_{24} +z_{24}^2}.
\end{equation}
Making use of these results, we were able to confirm {\it via} exact 
symbolic integrations, 
the (formally demonstrated) 
result of \.Zyczkowski and Sommers
\cite{szHS} that the HS volume of the {\it real} 
two-qubit ($n=4$) states is
$\frac{\pi^4}{60480} \approx 0.0016106$.
(This result was also achievable through a somewhat different
Mathematica computation, using the implicit integration feature
first 
introduced in version 5.1. That is, the only integration limits employed were
that $z_{ij} \in [-1,1], i \neq j$ --- {\it broader} than those in 
(\ref{limits}) --- while the Boolean constraints were imposed that 
the determinant of $\rho$ and {\it one} [all that is needed to ensure
nonnegativity] of its principal $3 \times 3$ minors be nonnegative.)
\subsection{Determinant of the Partial Transpose}
However, when we tried to combine these integration limits (\ref{limits}) 
with
the (Peres-Horodecki \cite{asher,michal,bruss} $n=4$) 
 separability constraint that the determinant ($C =|\rho_{PT}|$) 
of the partial
transpose of $\rho$ be nonnegative \cite[Thm. 5]{ver}, 
we exceeded the memory availabilities of our workstations.
In general, the term $C$ --- unlike the earlier term $B$ --- unavoidably 
involves 
the diagonal entries ($\rho_{ii}$), so the
dimension of the accompanying integration problems must increase, 
it would seem, we initially thought  --- 
in the $9$-dimensional real  $n=4$ case from 6 to 9.

However, we then noted that, in fact, the dimensionality 
of the required integrations must only
essentially be increased by one (rather than three), since 
$C$ turns out to be (aside from the necessarily nonnegative factor of $A$) 
expressible solely in terms of 
 the (six, in the real case) distinct
$z_{ij}$'s 
and the square root ratio
\begin{equation}
\mu = \sqrt{\frac{\rho_{11} \rho_{44}}{\rho_{22} \rho_{33}}}.
\end{equation}
(Considering $\mu$ as fixed, this is the equation of an hyperboloid of
one sheet \cite[p. 227]{CRC}.)
That is, 
\begin{equation}
C \equiv |\rho_{PT}| = 
 A  \Big(-z_{14}^2 \mu ^4+2 z_{14} \left(z_{12} z_{13}+z_{24}
   z_{34}\right) \mu ^3 + s \mu^2 
+2 z_{23}
   \left(z_{12} z_{24}+z_{13} z_{34}\right) \mu -z_{23}^2 \Big),
\end{equation}
where 
\begin{displaymath}
s= \left(z_{34}^2-1\right)
   z_{12}^2-2 \left(z_{14} z_{23}+z_{13} z_{24}\right)
   z_{34} z_{12}-z_{13}^2+z_{14}^2
   z_{23}^2+\left(z_{13}^2-1\right) z_{24}^2-z_{34}^2-2
   z_{13} z_{14} z_{23} z_{24}+1.
\end{displaymath}

$C$ is, thus,  a quartic/biquadratic polynomial in terms of $\mu$ 
(cf. \cite{wang,sudarshan}). 
(Clearly, the difficulty of the 
two-qubit separable volume problem under study here
is strongly tied to the high [fourth] degree of C in $\mu$. 
By setting either $z_{14}=0$ or $z_{23}=0$, the degree of $C$
can be reduced to 2 (cf. \cite{slaterJPAreject}).)
In the {\it complex} case, $C$ once again assumes the form of a
quartic polynomial in $\mu$. So one encounters, in that setting, 
thirteen-dimensional integration
problems rather than fifteen-dimensional ones.
\section{Analyses}
So, the problem of determining the separable volumes 
can be seen to hinge on (in the
real case), a {\it seven}-fold
integration involving the six (independent) $z_{ij}$'s and $\mu$.
However, such requisite integrations, allowing $\mu$ to vary (or even holding
$\mu$ constant at various values, thus, reducing to six-fold integrations), 
did not appear to be exactly/symbolically  
performable (using version 5.2 of 
Mathematica).
\subsection{Estimation of the univariate functions 
$f_{real}(\mu)$ and $f_{complex}(\mu)$}
Thus, to make further progress, it seemed necessary, at this stage, 
 to employ 
numerical methods (not excluding the possibility 
that exact solutions might, at some point,  be revealed).

We proceeded along two parallel courses, one for the 9-dimensional
real two-qubit case and the other for the 15-dimensional complex case.
We sought those functions $f_{real}(\mu)$ and $f_{complex}(\mu)$
that would result from imposing the conditions that 
the expressions $A$, $B$ and $C$
(as well as a principal $3 \times 3$ minor of $\rho$), along with
their complex counterpart expressions, 
be {\it simultaneously} 
nonnegative. (The satisfaction of these 
joint conditions ensures that we are dealing precisely
with {\it separable} $4 \times 4$ density matrices.) 
It was evident that the relation $f(\mu)=f(\frac{1}{\mu})$ must hold,
so we only studied the range $\mu \in [0,1]$. Dividing this unit interval
into 2,000 equal nonoverlapping subintervals of length $\frac{1}{2000}$ 
each, 
we sought to estimate the $f(\mu)$'s at the 2,001 end points of these
subintervals.

This required ($\mu$ being fixed at these end points) numerical
integrations in 6 and 12 dimensions. For this purpose, we utilized the
Tezuka-Faure (TF) quasi-Monte Carlo procedure \cite{giray1,tezuka}, 
we had extensively used in
our earlier studies of separability probabilities \cite{slaterJGP, slaterPRA}. 
For each of the 2,001 discrete,
equally-spaced values of $\mu$  we employed 
the same set of 37,000,000 Tezuka-Faure 
six-dimensional points in the real case and, similarly,  the 
 same
set of 25,000,000 twelve-dimensional points in the complex case.
(The Tezuka-Faure points are defined over unit hypercubes 
$[0,1]^{n}$, so in our computations, we transform the Bloore variables
accordingly. We plan to continue to add such points to our [real and complex]
analyses.)

\begin{figure}
\includegraphics{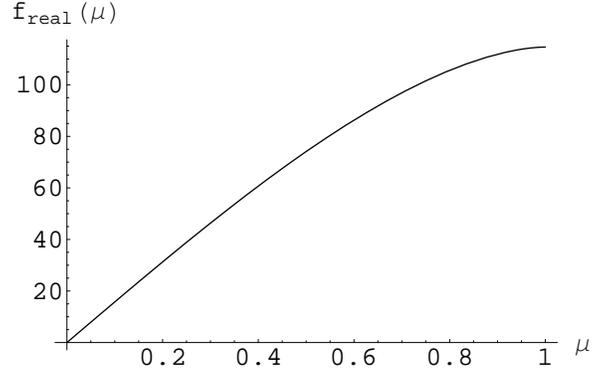}
\caption{\label{fig:freal}Estimation of $f_{real}(\mu)$ based on the 
third-order 
interpolation of 2,001 points ($\mu$), 
the value at each such point being based
on {\it six}-fold numerical integrations employing 
the same (for each $\mu$) set of thirty-seven million Tezuka-Faure points}
\end{figure}

\begin{figure}
\includegraphics{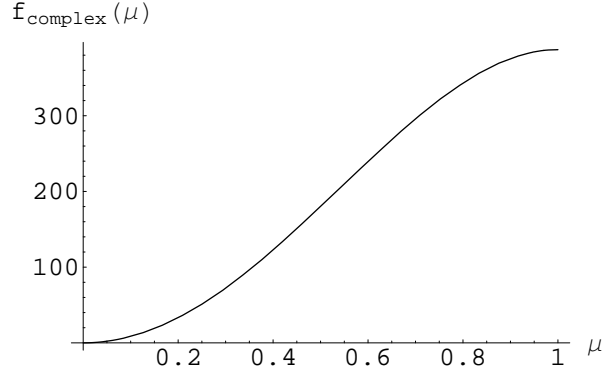}
\caption{\label{fig:fcomplex}Estimation of $f_{complex}(\mu)$ based on the
third-order
interpolation of 2,001 points ($\mu$), the value at each such point being based
on {\it twelve}-fold numerical integrations employing 
the same (for each $\mu$) set of twenty-five million Tezuka-Faure points}
\end{figure}

In Figs.~\ref{fig:freal} and \ref{fig:fcomplex} we show the results
of this procedure. 
There were some slight deviations from monotonicity \cite{ramsay}
(presumably due to limited sample sizes) in the vicinity
of $\mu=1$ for both functions.

In the real case, our {\it estimate} of
{\it known} Hilbert-Schmidt volume of (separable {\it plus} 
nonseparable) states \cite{szHS}, 
$\frac{\pi^4}{60480} \approx 0.0016106$ 
was larger by only a factor 1.00006. So, we would expect our 
companion estimates
of $f_{real}(\mu)$, at each of the 2,001 sampled points, 
to be roughly equally precise. 
(Let us note that $f_{real}(0) = f_{complex}(0)=0$.) 
In the complex case,
our estimate of the known
15-dimensional volume, $\frac{\pi^6}{851350500} \approx 
1.12925 \cdot 10^{-6}$ was smaller only by a factor of 0.99965. 
(As instances of specific values, based on independent analyses using
still larger numbers of TF-points,
we obtained estimates of $f_{real}(1) = \frac{73430796}{640625} \approx 
114.62368, 
f_{real}(\frac{1}{2}) =  \frac{47475904}{640625} \approx 
74.108728, f_{real}(\sigma_{Au}) = 
\frac{56575096}{640625} \approx 88.312344, f_{complex}(1) = 387.33307366, 
f_{complex}(\frac{1}{2}) = 180.6046580, f_{complex}(\sigma_{Au}) = 
251.157815860$, where $\sigma_{Au} =\frac{\sqrt{5}-1}{2}$ 
denotes the golden ratio \cite{livio}. {\it Exact} characterizations 
of $f_{real}(\mu)$ and $f_{complex}(\mu)$ would, of course, be of
great interest, in particular, for the possibility that they might
yield exact volume results.)

To estimate the desired separable volumes ($V_{sep}$) themselves, one must
perform the calculations,
\begin{equation} \label{Vreal}
V_{sep/real} = 2 \int_{0}^{1} jac_{real} f_{real}(\mu) d \mu
\end{equation} 
and 
\begin{equation} \label{Vcomplex}
V_{sep/complex} = 2 \int_{0}^{1} jac_{complex} f_{complex}(\mu) d \mu.
\end{equation}
\subsection{Jacobians for the transformations}
\begin{figure}
\includegraphics{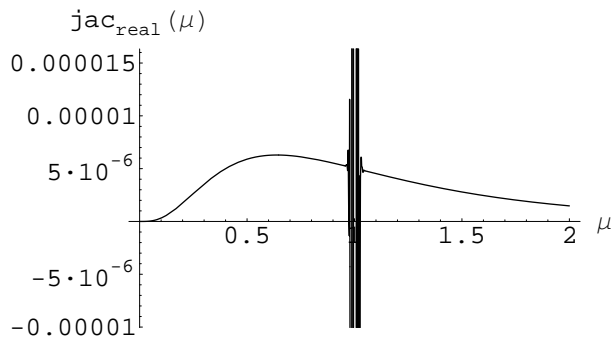}
\caption{\label{fig:jacreal}Plot of the jacobian function
$jac_{real}(\mu)$, given by (\ref{jacr})}
\end{figure}

\begin{figure}
\includegraphics{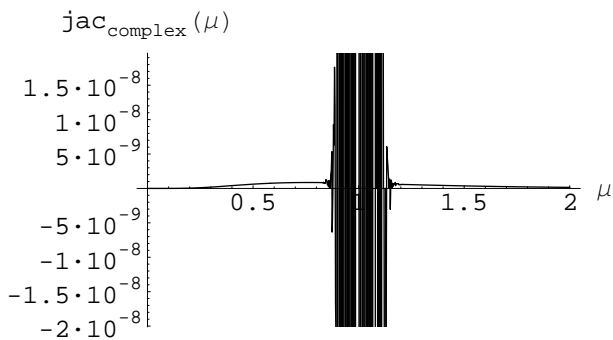}
\caption{\label{fig:jaccomplex}Plot of the jacobian function
$jac_{complex}(\mu)$, given by (\ref{jacc})}
\end{figure}

Now
(Fig.~\ref{fig:jacreal}),
\begin{equation} \label{jacr}
jac_{real}(\mu)=
\frac{\mu ^4 \left(12 \left(\left(\mu ^2+2\right)
   \left(\mu ^4+14 \mu ^2+8\right) \mu ^2+1\right) \log
   (\mu )-5 \left(5 \mu ^8+32 \mu ^6-32 \mu
   ^2-5\right)\right)}{1890 \left(\mu ^2-1\right)^9}
\end{equation}
and (Fig.~\ref{fig:jaccomplex})
\begin{equation} \label{jacc}
jac_{complex}(\mu)= -\frac{\mu ^7}{1801800 \left(\mu ^2-1\right)^{15}} V,
\end{equation} 
where
\begin{displaymath}
V =363 \mu ^{14}+9947 \mu ^{12}+48363 \mu ^{10}+42875 \mu
   ^8-42875 \mu ^6-48363 \mu ^4-9947 \mu ^2-363
\end{displaymath}
\begin{displaymath}
-140 \left(\mu ^{14}+49 \mu ^{12}+441 \mu ^{10}+1225 \mu
   ^8+1225 \mu ^6+441 \mu ^4+49 \mu ^2+1\right) \log (\mu
   ).
\end{displaymath}
We have that 
\begin{equation}
\int_{0}^{1} jac_{real}(\mu) d 
\mu = \frac{\pi ^2}{2293760} \approx 4.30281 \cdot 
10^{-6},
\end{equation}
and
\begin{equation}
  \int_{0}^{1} jac_{complex}(\mu) d \mu = \frac{1}{2018016000} 
\approx 4.95536 \cdot 10^{-10}.
\end{equation}
(The smallest value of $\mu$ for which $jac_{real}(\mu) =0$ that we were 
able to find was 0.9685588023, while for $jac_{complex}(\mu)$ we found  
0.8395384257.) 

We obtained the jacobian functions $jac_{real}(\mu)$ and
$jac_{complex}(\mu)$, given in (\ref{jacr}) and (\ref{jacc}),
by transformations of, say, $\rho_{33}$ to the $\mu$ variable 
(and subsequent two-fold exact integrations over $\rho_{11}$ and 
$\rho_{22}$) of  
the original (three-dimensional) jacobians, 
involving the diagonal entries, for the Bloore parameterizations. These 
original 
jacobians were of the form 
$(\Pi_{i=1}^{4} \rho_{ii})^k$
with
$k=\frac{3}{2}$ in the real case, and $k=3$, in the complex case.
(Of course, by the unit trace condition, 
we must have $\rho_{44}=1-\rho_{11}-\rho_{22}-\rho_{33}$.)

The direct high-accuracy computation of the desired 
separable volume integrals (\ref{Vreal}) and
(\ref{Vcomplex}) proves challenging due to the highly oscillatory nature
of $jac_{real}(\mu)$ and $jac_{complex}(\mu)$ (given by 
(\ref{jacr})  and 
(\ref{jacc})) in the vicinity of 
$\mu=1$, as indicated in Figs.~\ref{fig:jacreal} and \ref{fig:jaccomplex}. 
(It might be appropriate to sample more points in the vicinity of
$\mu=1$ than in other less problematical regions. 
We have also  attempted --- without significant success so far --- to 
evaluate these integrals using repeated integration
by parts \cite{manning}, since the two jacobians in question 
admit repeated exact integrations.)
\subsection{Volume integrals over $\mu \in [0,.95]$}
Replacing the upper integration limit of 1 in the integral (\ref{Vreal})
by .95, we obtained --- using high precision arithmetic --- a 
result of 0.0006707668  and 
consequent {\it lower} bound
on the probability of separability of the real two-qubit systems of
0.41647013. (The direct 
use of upper integration limits greater than .95
appeared to lead to unstable results.)
For similar reasons, replacing the upper integration limit of 1 
in the integral (\ref{Vcomplex}) also by .95, we obtained a result of
$2.327058044  \cdot 10^{-7}$ 
and consequent lower bound on the probability of separability of the
complex two-qubit systems of 0.2060707612.
\subsection{Volume integrals over $\mu \in [.95,1]$} \label{mike}
\begin{figure}
\includegraphics{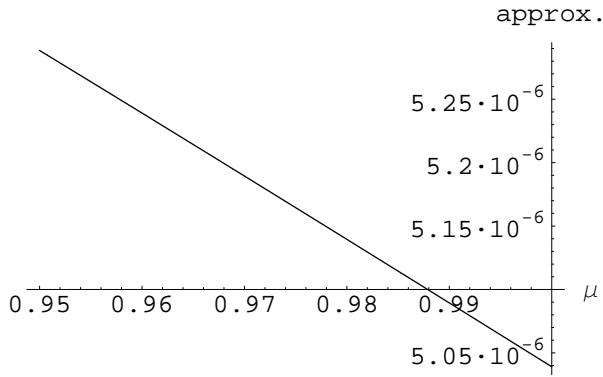}
\caption{\label{fig:SeriesReal}100-degree power series approximation to
$jac_{real}(\mu)$ about $\mu=1$}
\end{figure}

\begin{figure}
\includegraphics{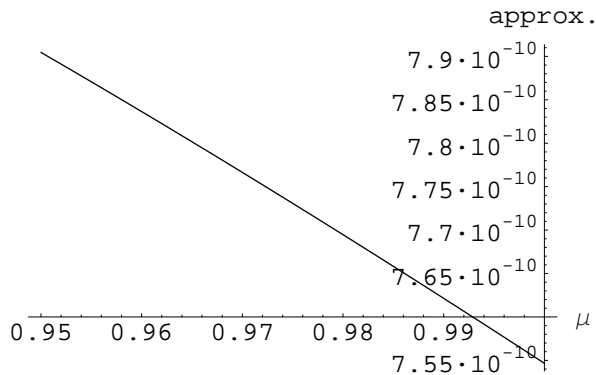}
\caption{\label{fig:SeriesComplex}100-degree power series approximation  to
$jac_{complex}(\mu)$ about $\mu=1$}
\end{figure}

In the immediately preceding analysis, we used upper limits of .95 
rather than 1 in the integrals (\ref{Vreal}) and (\ref{Vcomplex}).
To estimate the integrals in the remaining range $[.95,1]$, we replaced
the jacobian functions $jac_{real}(\mu)$  and $jac_{complex}(\mu)$, given in
(\ref{jacr}) and (\ref{jacc}), by their 100-degree power 
series expansions about $\mu=1$. (When plotted over [.95,1], both these
replacement functions gave the appearances of simple downward-sloping
{\it lines} (Figs.~\ref{fig:SeriesReal} and \ref{fig:SeriesComplex}).)

Proceeding in such a manner, again using high-precision arithmetic 
and summing the results over the two sets of intervals, 
we arrived at our final (subject to the availability of additional 
Tezuka-Faure points) estimates
$V_{sep/real} \approx 0.0007298112$ and 
$V_{sep/complex} \approx 2.625622678 \cdot 10^{-7}$. 
Then, we have
$prob_{sep/real} \approx 0.45313001$ and
$prob_{complex/real} \approx  0.23250991$.
(When we compared these several results, based on interpolation --- to 
estimate $f_{real}(\mu)$ and $f_{complex}(\mu)$ (Figs.~\ref{fig:freal} 
and \ref{fig:fcomplex}) --- using
{\it third}-degree polynomials with those using {\it sixth}-degree 
polynomials, we 
obtained essentially the same set of results.)

As noted in the introductory section,
we had previously hypothesized that $V_{sep/complex}= 
(5 \sqrt{3})^{-7} \approx 2.73707 \cdot 10^{-7}$ 
\cite[eq. (41)]{slaterPRA} and
$prob_{sep/complex}=  \frac{2^2 \cdot 3 \cdot 7^2 \cdot 11 
\cdot 13 \sqrt{3}}{5^4 \pi^{6}} 
\approx 0.242379$ \cite[eq. (43), but misprinted as $5^3$ not
$5^4$ there]{slaterPRA}. The analysis there was based on a considerably
larger number --- 400,000,000 --- of Tezuka-Faure points than here.
But each point there was employed only {\it once} for the Peres-Horodecki
separability test, while each point here is used in 2,000 such tests
(with $\mu$ ranging over [0,1]).
(Relatedly, we had initially suspected that if we started 
checking the Peres-Horodecki criterion for successively larger values 
of  $\mu$, holding the set of $z_{ij}$'s given by a Tezuka-Faure point 
{\it fixed},  then if we reached one value for which separability held, then
{\it all} higher values of $\mu$ would also yield separability. But this turned out
not to be invariably the case. So, it appeared that 
we needed to check the criterion 2,000 times
for each point.)

\section{Concluding Remarks}
In our earlier study \cite{slaterJPAreject}, we had also employed
the Bloore parameterization of the 
two-qubit (and qubit-qutrit) systems
to study the Hilbert-Schmidt separability probabilities of 
specialized systems of
less than full dimensionality.
We also reported an effort to determine a certain {\it three}-dimensional
function (in contrast to the {\it one}-dimensional functions 
$f_{real}(\mu)$ and $f_{complex}(\mu)$ above, but for somewhat a similar
purpose) 
over the simplex of eigenvalues that would facilitate the
calculation of the 15-dimensional volume of the two-qubit systems in terms
of (monotone) metrics --- such as the Bures, Kubo-Mori, 
Wigner-Yanase,\ldots --- other than the (non-monotone 
\cite{ozawa}) Hilbert-Schmidt one considered here.
(The Bloore parameterization \cite{bloore}, used above, did not seem
immediately useful in this monotone metric context, since the {\it 
eigenvalues} of
$\rho$ are not explicitly expressed (cf. \cite{Dittmann}). 
Therefore, we had recourse in \cite{slaterJPAreject} to
the Euler-angle parameterization of Tilma, Byrd and Sudarshan
\cite{sudarshan}.)

Let us direct the reader to some papers of R. Kellerhals concerned, among
other items, with the volumes of {\it hyperbolic} polyhedra 
\cite{ruth1,ruth2} (cf. \cite{freitas}).
In this line of work, the dilogarithm and, more generally, the polylogarithm
functions play important roles. There have been some indicators in our
investigations above (in particular, in integrations of the jacobians) 
that these functions also may be of relevance in our context.

The extension to qubit-{\it qutrit} pairs (and even higher-dimensional compositie systems) of the univariate-function-strategy we have pursued above,
for the case of qubit-qubit pairs, seems problematical, although we have not
yet examined the matter in 
great detail. In the qubit-qubit case, the analysis is
facilitated by the fact that it is sufficient that the determinant of
the partial transpose be nonnegative for the Peres-Horodecki separability
criterion to hold \cite[Thm. 5]{ver}. 
More requirements than this single one
are needed in the qubit-qutrit
scenario --- even though the criterion on the nonnegativity of the 
partial transpose  is still both necessary and 
sufficient for $6 \times 6$ density matrices. 
(In addition to the determinant, the 
leading minors and/or
the individual eigenvalues of the partial transpose of the 
$6 \times 6$ density matrix 
would need to be tested for nonnegativity, as well. 
Also the qubit-qutrit analogue
of the ratio ($\mu$) of diagonal entries would have to be defined, if
even possible.)

\begin{acknowledgments}
I wish to express gratitude to the Kavli Institute for Theoretical
Physics (KITP)
for computational support in this research and to Michael Trott for his
suggestion regarding the use of the  power series expansion 
in Sec.~\ref{mike}.

\end{acknowledgments}

\bibliography{SV1}

\end{document}